\documentclass[%
 reprint,
 amsmath,amssymb,
 aps,
]{revtex4-2}

\usepackage{graphicx}
\usepackage{dcolumn}
\usepackage{bm}
\usepackage{hyperref}
\usepackage{xspace}
\usepackage{subcaption}

\begin{document}

%

\newcommand{\pp}           {$pp$\xspace}
\newcommand{\ppbar}        {\mbox{$\mathrm {p\overline{p}}$}\xspace}
\newcommand{\XeXe}         {\mbox{Xe--Xe}\xspace}
\newcommand{\PbPb}         {\mbox{Pb--Pb}\xspace}
\newcommand{\pA}           {\mbox{pA}\xspace}
\newcommand{\pPb}          {\mbox{p--Pb}\xspace}
\newcommand{\AuAu}         {\mbox{Au--Au}\xspace}
\newcommand{\dAu}          {\mbox{d--Au}\xspace}

\newcommand{\s}            {\ensuremath{\sqrt{s}}\xspace}
\newcommand{\snn}          {\ensuremath{\sqrt{s_{\mathrm{NN}}}}\xspace}
\newcommand{\pt}           {\ensuremath{p_{\rm T}}\xspace}
\newcommand{\pT}           {\ensuremath{p_\mathrm{T}}\xspace}
\newcommand{\pTchjet}      {\ensuremath{p_\mathrm{T,ch\,jet}}\xspace}
\newcommand{\kT}           {\ensuremath{k_\mathrm{T}}\xspace}
\newcommand{\sqsn}         {\ensuremath{\sqrt{s_\mathrm{NN}}}\xspace}
\newcommand{\sqs}          {\ensuremath{\sqrt{s}}\xspace}

\newcommand{\meanpt}       {$\langle p_{\mathrm{T}}\rangle$\xspace}
\newcommand{\ycms}         {\ensuremath{y_{\rm CMS}}\xspace}
\newcommand{\ylab}         {\ensuremath{y_{\rm lab}}\xspace}
\newcommand{\etarange}[1]  {\mbox{$\left | \eta \right |~<~#1$}}
\newcommand{\yrange}[1]    {\mbox{$\left | y \right |~<~#1$}}
\newcommand{\dndy}         {\ensuremath{\mathrm{d}N_\mathrm{ch}/\mathrm{d}y}\xspace}
\newcommand{\dndeta}       {\ensuremath{\mathrm{d}N_\mathrm{ch}/\mathrm{d}\eta}\xspace}
\newcommand{\dndptBjet}    {\ensuremath{\mathrm{d}N_\mathrm{b-jet}/\mathrm{d}\pt}\xspace}
\newcommand{\avdndeta}     {\ensuremath{\langle\dndeta\rangle}\xspace}
\newcommand{\dNdy}         {\ensuremath{\mathrm{d}N_\mathrm{ch}/\mathrm{d}y}\xspace}
\newcommand{\Npart}        {\ensuremath{N_\mathrm{part}}\xspace}
\newcommand{\Ncoll}        {\ensuremath{N_\mathrm{coll}}\xspace}
\newcommand{\dEdx}         {\ensuremath{\textrm{d}E/\textrm{d}x}\xspace}
\newcommand{\RpPb}         {\ensuremath{R_{\rm pPb}}\xspace}
\newcommand{\RpPbBjet}     {\ensuremath{R_{\rm pPb}^\text{b-jet}}\xspace}
\newcommand{\RAA}          {\ensuremath{R_{\rm AA}}\xspace}

\newcommand{\nineH}        {$\sqrt{s}~=~0.9$~Te\kern-.1emV\xspace}
\newcommand{\seven}        {$\sqrt{s}~=~7$~Te\kern-.1emV\xspace}
\newcommand{\twoH}         {$\sqrt{s}~=~0.2$~Te\kern-.1emV\xspace}
\newcommand{\twosevensix}  {$\sqrt{s}~=~2.76$~Te\kern-.1emV\xspace}
\newcommand{\five}         {$\sqrt{s}~=~5.02$~Te\kern-.1emV\xspace}
\newcommand{\twosevensixnn}{$\sqrt{s_{\mathrm{NN}}}~=~2.76$~Te\kern-.1emV\xspace}
\newcommand{\fivenn}       {$\sqrt{s_{\mathrm{NN}}}~=~5.02$~Te\kern-.1emV\xspace}
\newcommand{\LT}           {L{\'e}vy-Tsallis\xspace}
\newcommand{\GeVc}         {Ge\kern-.1emV/$c$\xspace}
\newcommand{\MeVc}         {Me\kern-.1emV/$c$\xspace}
\newcommand{\TeV}          {Te\kern-.1emV\xspace}
\newcommand{\GeV}          {Ge\kern-.1emV\xspace}
\newcommand{\MeV}          {Me\kern-.1emV\xspace}
\newcommand{\GeVmass}      {Ge\kern-.2emV/$c^2$\xspace}
\newcommand{\MeVmass}      {Me\kern-.2emV/$c^2$\xspace}
\newcommand{\lumi}         {\ensuremath{\mathcal{L}}\xspace}

\newcommand{\ITS}          {\rm{ITS}\xspace}
\newcommand{\TOF}          {\rm{TOF}\xspace}
\newcommand{\ZDC}          {\rm{ZDC}\xspace}
\newcommand{\ZDCs}         {\rm{ZDCs}\xspace}
\newcommand{\ZNA}          {\rm{ZNA}\xspace}
\newcommand{\ZNC}          {\rm{ZNC}\xspace}
\newcommand{\SPD}          {\rm{SPD}\xspace}
\newcommand{\SDD}          {\rm{SDD}\xspace}
\newcommand{\SSD}          {\rm{SSD}\xspace}
\newcommand{\TPC}          {\rm{TPC}\xspace}
\newcommand{\TRD}          {\rm{TRD}\xspace}
\newcommand{\VZERO}        {\rm{V0}\xspace}
\newcommand{\VZEROA}       {\rm{V0A}\xspace}
\newcommand{\VZEROC}       {\rm{V0C}\xspace}
\newcommand{\Vdecay} 	   {\ensuremath{V^{0}}\xspace}

\newcommand{\ee}           {\ensuremath{e^{+}e^{-}}} 
\newcommand{\pip}          {\ensuremath{\pi^{+}}\xspace}
\newcommand{\pim}          {\ensuremath{\pi^{-}}\xspace}
\newcommand{\kap}          {\ensuremath{\rm{K}^{+}}\xspace}
\newcommand{\kam}          {\ensuremath{\rm{K}^{-}}\xspace}
\newcommand{\pbar}         {\ensuremath{\rm\overline{p}}\xspace}
\newcommand{\kzero}        {\ensuremath{{\rm K}^{0}_{\rm{S}}}\xspace}
\newcommand{\lmb}          {\ensuremath{\Lambda}\xspace}
\newcommand{\almb}         {\ensuremath{\overline{\Lambda}}\xspace}
\newcommand{\Om}           {\ensuremath{\Omega^-}\xspace}
\newcommand{\Mo}           {\ensuremath{\overline{\Omega}^+}\xspace}
\newcommand{\X}            {\ensuremath{\Xi^-}\xspace}
\newcommand{\Ix}           {\ensuremath{\overline{\Xi}^+}\xspace}
\newcommand{\Xis}          {\ensuremath{\Xi^{\pm}}\xspace}
\newcommand{\Oms}          {\ensuremath{\Omega^{\pm}}\xspace}
\newcommand{\degree}       {\ensuremath{^{\rm o}}\xspace}

\title{Machine learning-based b-jet tagging in \pp collisions at  \s = 13 TeV}

\author{Hadi Hassan$^{1}$}
\email[Contact author: ]{hadi.hassan@cern.ch}

\author{Neelkamal Mallick$^{2}$}
\email[Contact author: ]{neelkamal.mallick@cern.ch}


\author{~D.J.~Kim$^{2,3}$}%
\email[Contact author: ]{dong.jo.kim@cern.ch}
\affiliation{$^{1}$University of Tsukuba, Tennodai 1-1-1, Tsukuba, Ibaraki 305-0006, Japan}
\affiliation{$^{2}$University of Jyväskylä, Department of Physics, P.O. Box 35, FI-40014, Jyväskylä, Finland}
\affiliation{$^{3}$Helsinki Institute of Physics, P.O.Box 64, FI-00014, Helsinki, Finland}

\date{\today}

\begin{abstract}

Studying heavy-flavor jets in \pp collisions is important since it can test perturbative QCD calculations and be used as a reference for heavy-ion collisions. Jets in this analysis are reconstructed from charged particles using the anti-$k_{\mathrm{T}}$ algorithm with a resolution parameter $R=$ 0.4 at midrapidity $|\eta|<$ 0.5. Beauty jets are tagged using a machine learning model that uses a convolutional neural network trained on information extracted from the jet, tracks, and secondary vertices. Results show that this model is superior in b-jet-tagging compared to other traditional tagging methods.

\end{abstract}

\maketitle


\section{Introduction}\label{sec:Intro}

Heavy-flavor jets are important for the understanding of quantum chromodynamic (QCD)~\cite{Politzer:1974fr}. They can provide tests for the perturbative QCD calculations, and have a large mass that acts as a hard scale which suppresses soft-gluon radiations at small angles~\cite{Dokshitzer:1991fd,ALICE:2021aqk}. They are produced very early in the collision, so they carry information on the hard collision and information about the high-energy-density medium produced in heavy-ion collisions~\cite{Zhang:2003wk,Dokshitzer:2001zm}. Identification of jets formed from hadronization of b-quarks, so-called ``b-jets", is important for measurements such as cross section~\cite{CMS:2012pgw,ALICE:2021wct,CMS:2021pcj}, fragmentation, jet substructure~\cite{CMS:2020geg,CMS:2022btc}, and for understanding the quark-gluon plasma in heavy-ion collisions~\cite{ATLAS:2022agz,CMS:2013qak}.

Various tagging algorithms have been developed to achieve the heavy-flavor jet identification. The problem with traditional b-jet tagging algorithms, such as impact parameter (IP) or secondary vertex (SV) methods~\cite{ALICE:2021wct,CMS:2012feb,ATL-PHYS-PUB-2017-011,ATL-PHYS-PUB-2017-013} that exploit the long lifetime of the b-hadron, is that they rely on cuts imposed on a single variable. Machine learning (ML) methods can combine several parameters from different reconstruction levels, such as track-level observables (e.g., track impact parameters), jet-level properties (e.g., jet mass), or secondary vertex features (e.g., decay length), and extract complex multidimensional correlations and patterns among the event characteristics. Previous studies from ATLAS and CMS experiments show that the ML models are superior compared to the traditional methods in terms of performance~\cite{ATLAS:2023gog,ATLAS:2022qxm,app122010574,Bols:2020bkb}.
In this study, to simulate b-jet events in \pp collisions, the PYTHIA8 Monte Carlo (MC) event generator is used~\cite{pythia8,Bierlich:2022pfr}, mainly since it is widely used and carefully validated by the LHC experiments~\cite{ALICE:2020swj,CMS:2019csb}.
Experimental measurements, such as those performed by ALICE, are affected by detector effects such as finite tracking, pointing resolutions, acceptance, and efficiency. Although the analysis methods can be optimized for experimental conditions such as the ALICE environment, we wish to keep the discussion at the methodological level. Hence, we do not perform a complete MC simulation including the ALICE detector~\cite{ALICE} using GEANT4~\cite{AGOSTINELLI2003250}, but instead smear the transverse momentum (\pt) and distance to the closest approach (DCA) distributions to mimic realistic experimental conditions.

This paper is organized as follows: Sec.~\ref{sec:method} describes the event generation, spectra smearing, jet reconstruction, feature extraction, and machine learning methodology. Section~\ref{sec:results} presents the performance of the ML algorithm for the tagging of beauty jets and compares it with traditional methods such as the impact parameter and the secondary vertex methods. Finally, Sec.~\ref{sec:conclusion} summarizes the findings of this analysis.

\section{Methodology}\label{sec:method}
\subsection{Event generation and jet reconstruction}\label{sec:Events}
Proton-proton (\pp) collisions at center-of-mass energy \s~=~13 TeV are generated using the PYTHIA8~\cite{pythia8,Bierlich:2022pfr} MC event generator with the Monash 2013 tune~\cite{Skands:2014pea} and all the hard QCD processes. 
For a more realistic description of the ALICE detector~\cite{ALICE}, events with primary vertices within $|z| < 10$~cm in the beam direction, and final state charged particles (after parton shower, hadronization and decays) with transverse momentum \pt $>$ 0.15 \GeVc, within the ALICE tracker acceptance $| \eta |\, <$ 0.9, are used in this analysis. 

Then a transverse momentum smearing is applied on these tracks based on the parameters from the ALICE inner tracking system 2 (ITS2) momentum resolution~\cite{Abelev:1625842} using the following equation:
\begin{equation}
    \sigma(\pt)/\pt = \sqrt{a^2 + b^2/\pt^2 + c^2\cdot \pt^2}. 
\end{equation}
Here, $a=$ 0.005, $b=$ $8\times10^{-4}$ \GeVc, and $c=$ 0.001 (\GeVc)$^{-1}$ are constants that represents the momentum resolution of the ALICE ITS2 detector. In addition to the \pt smearing, a smearing of the DCA of the particles with respect to the primary vertex (PV) has been applied according to the DCA resolution of the ITS2 detector~\cite{Abelev:1625842, Triolo_2025},
\begin{equation}
    \sigma({\rm DCA}_{xy/z})\, (\mu \mathrm{m}) = \sqrt{a^2 + b^2/\pt^2}.
\end{equation}
Here, $a = 7$ and $6~\mu\mathrm{m}$, and $b = 23$ and $25~\mu\mathrm{m}\cdot\mathrm{GeV}/c$ are the  values of the constants for the DCA resolution in the transverse plane and $z$ direction, respectively. 

The selected particles with \pt and DCA smearing are passed to the jet clustering algorithm. Jets are reconstructed using the anti-$k_{\mathrm{T}}$ jet clustering algorithm~\cite{Cacciari:2008gp} from the FastJet package~\cite{Cacciari:2011ma} with a resolution parameter $R=$ 0.4. This ensures that most of the initial partonic energy is contained inside the clustered jet. The energies of the tracks are combined using the energy-recombination scheme ($E$-scheme), which includes the mass of the particles in the jet four-momentum calculation with the best strategy. It should be noted here that the real masses of the particles are used for the jet reconstruction. Further, a reconstructed jet is classified as a beauty jet (b-jet) if it contains a beauty quark (from the PYTHIA particle record), a charm jet (c-jet) if it contains a charm quark, and a light-flavor jet (lf-jet) if neither of the heavy quarks are present. Jets with beauty quarks coming from the gluon splitting are classified as b-jets.

\subsection{Feature extraction}\label{sec:features}
\begin{figure}[htb]
    \centering
    \subfloat{
        \includegraphics[width=0.7\linewidth]{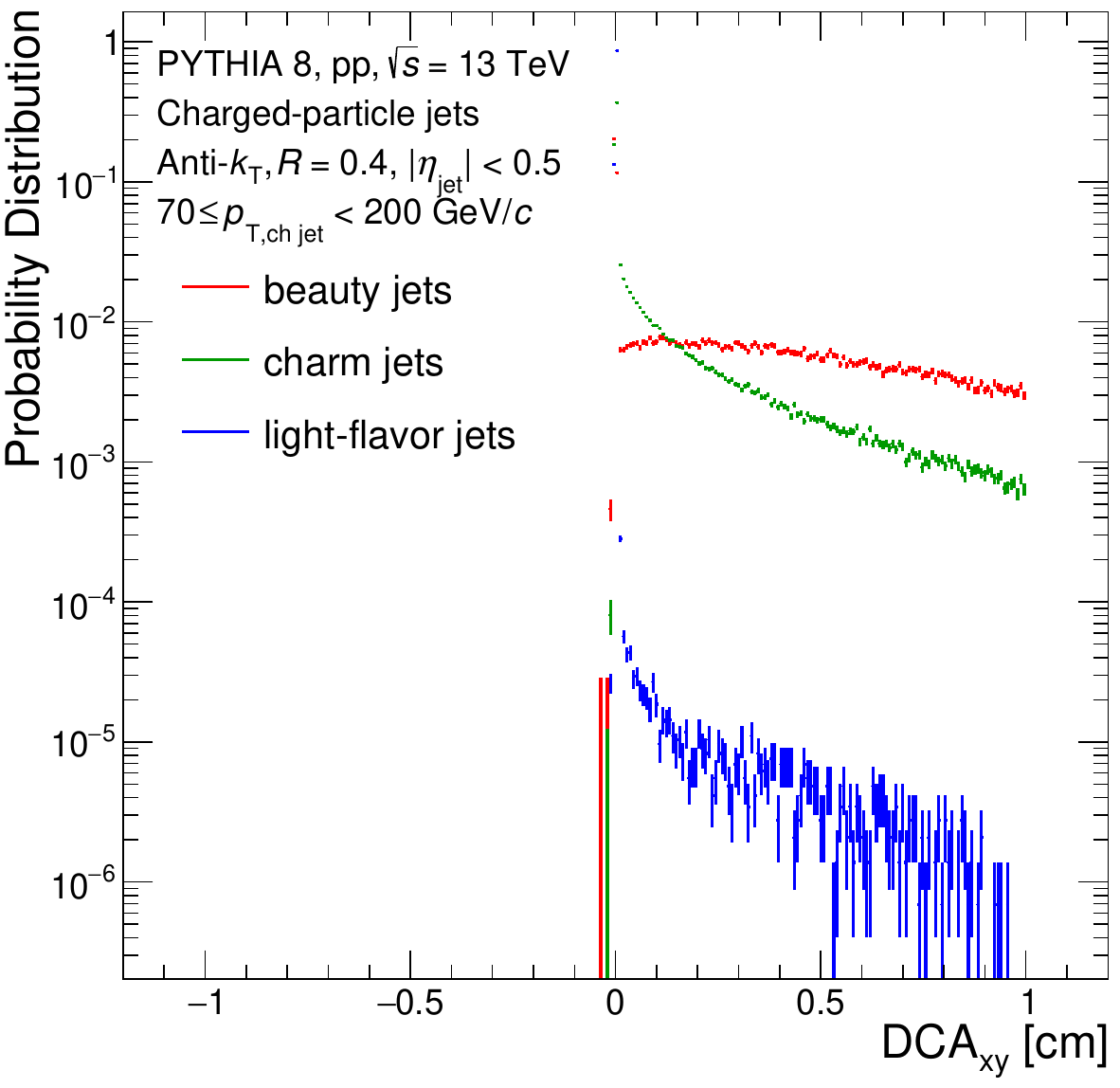}
        \label{fig:DCAxy}
    }
    \hfill
    \subfloat{
        \includegraphics[width=0.7\linewidth]{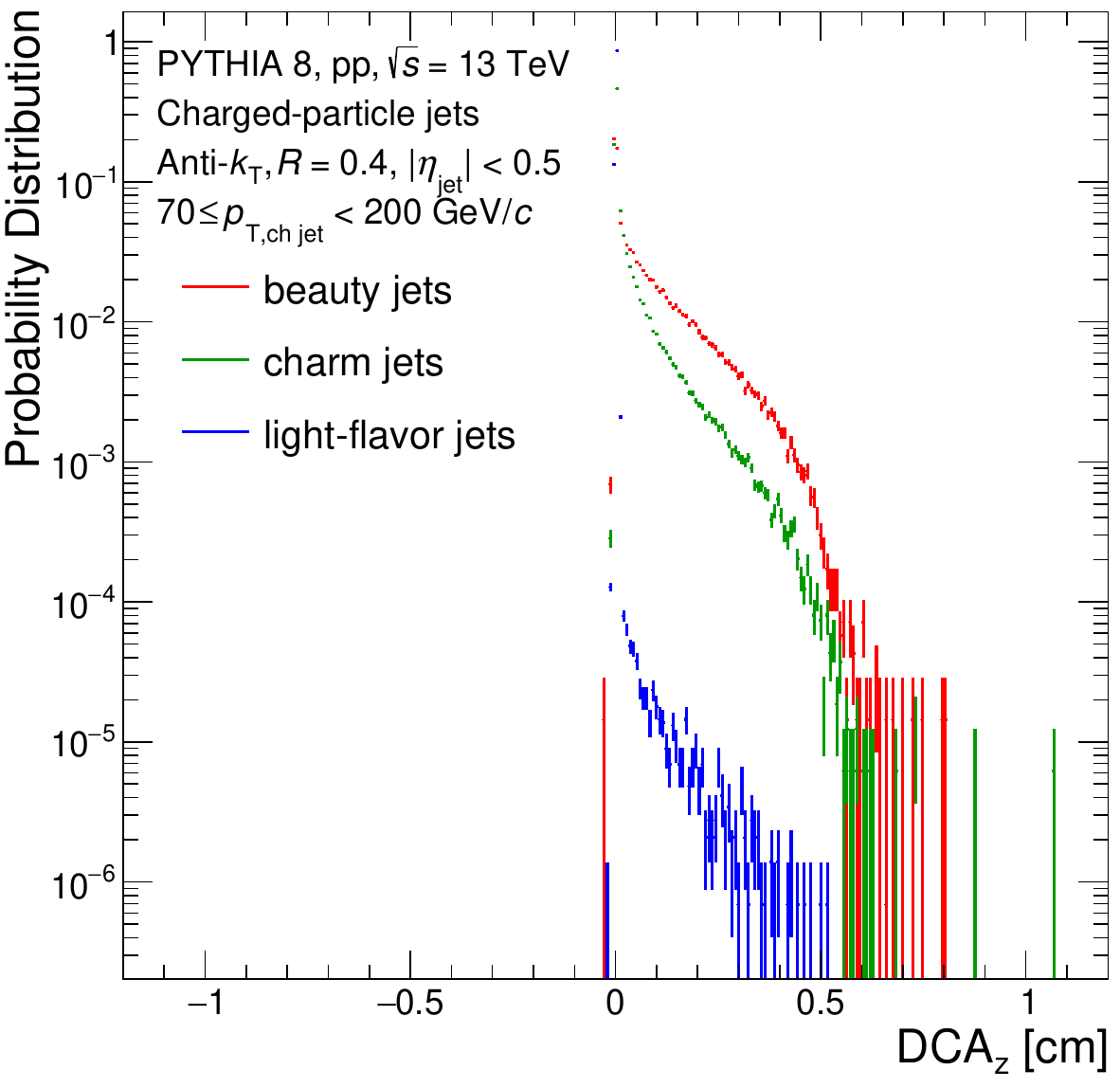}
        \label{fig:DCAz}
    }
    \caption{Transverse (top) and longitudinal (bottom) DCA distributions for different jet flavors.}
    \label{fig:DCA}
\end{figure}

For the machine learning algorithm, several features are extracted from the reconstructed jet, tracks, and secondary vertices.
For jets, the extracted features are the jet transverse momentum (\pt), pseudorapidity ($\eta$), azimuthal angle ($\varphi$), number of track constituents, number of reconstructed secondary vertices (explained later), and the jet mass. For the constituents of the jet, the extracted features are particle \pt, particle $\eta$, the dot product of the momenta of the track and jet ($\vec{p}_{\rm jet} \cdot \vec{p}_{\rm part}$), the dot product of the track and the jet divided by the jet momentum ($\vec{p}_{\rm jet} \cdot \vec{p}_{\rm part}/|\vec{p}_{\rm jet}|$), and the distance between the track and the jet axis in the $\eta-\varphi$ plane calculated as $\Delta R = \sqrt{(\eta_{\mathrm{jet}} - \eta_{\mathrm{track}})^2 - (\varphi_{\mathrm{jet}} - \varphi_{\mathrm{track}})^2}$. In addition, other features such as the signed DCA in the transverse ($xy$) plane, the signed DCA in $z$ direction, and the distance between the track and the closest reconstructed secondary vertex are calculated with the same distance equation mentioned before. The DCA sign is calculated as the sign of the dot product of the jet axis and the DCA vector ($\mathrm{sign}(\vec{\rm DCA}_{\rm track} \cdot \vec{p}_{\rm jet})$). Tracks coming from secondary vertices have positive DCA signs due to their long flight distance, while primary tracks coming from the primary vertex could have positive or negative DCA signs. Some of these features extracted from the jet and tracks are very discriminative between the different jet flavors, especially the signed DCA of the track to the primary vertex, as seen in Fig~\ref{fig:DCA}.

Prior to particle selection, a maximum DCA (in both $xy$ and $z$) of 1 cm is applied to reject tracks originating from strange particle decays. Since the DCA is the most discriminative observable, the tracks in each jet are sorted in descending order of their signed DCAs, and the top ten tracks are selected as input to the ML model.

\begin{figure*}[htb]
    \centering
    \subfloat{
        \includegraphics[width=0.45\linewidth]{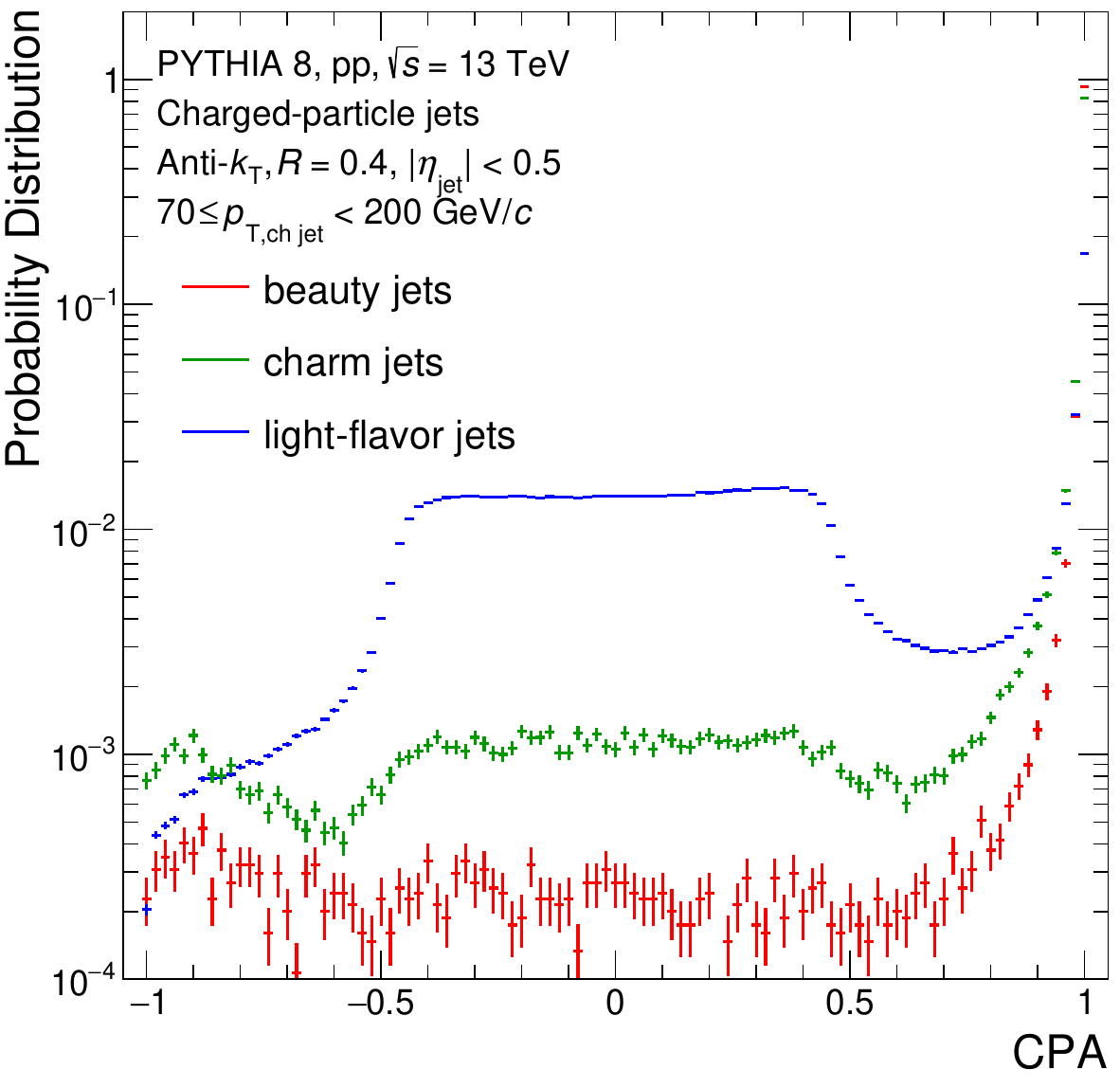}
        \label{fig:CPA}
    }
    \subfloat{
        \includegraphics[width=0.45\linewidth]{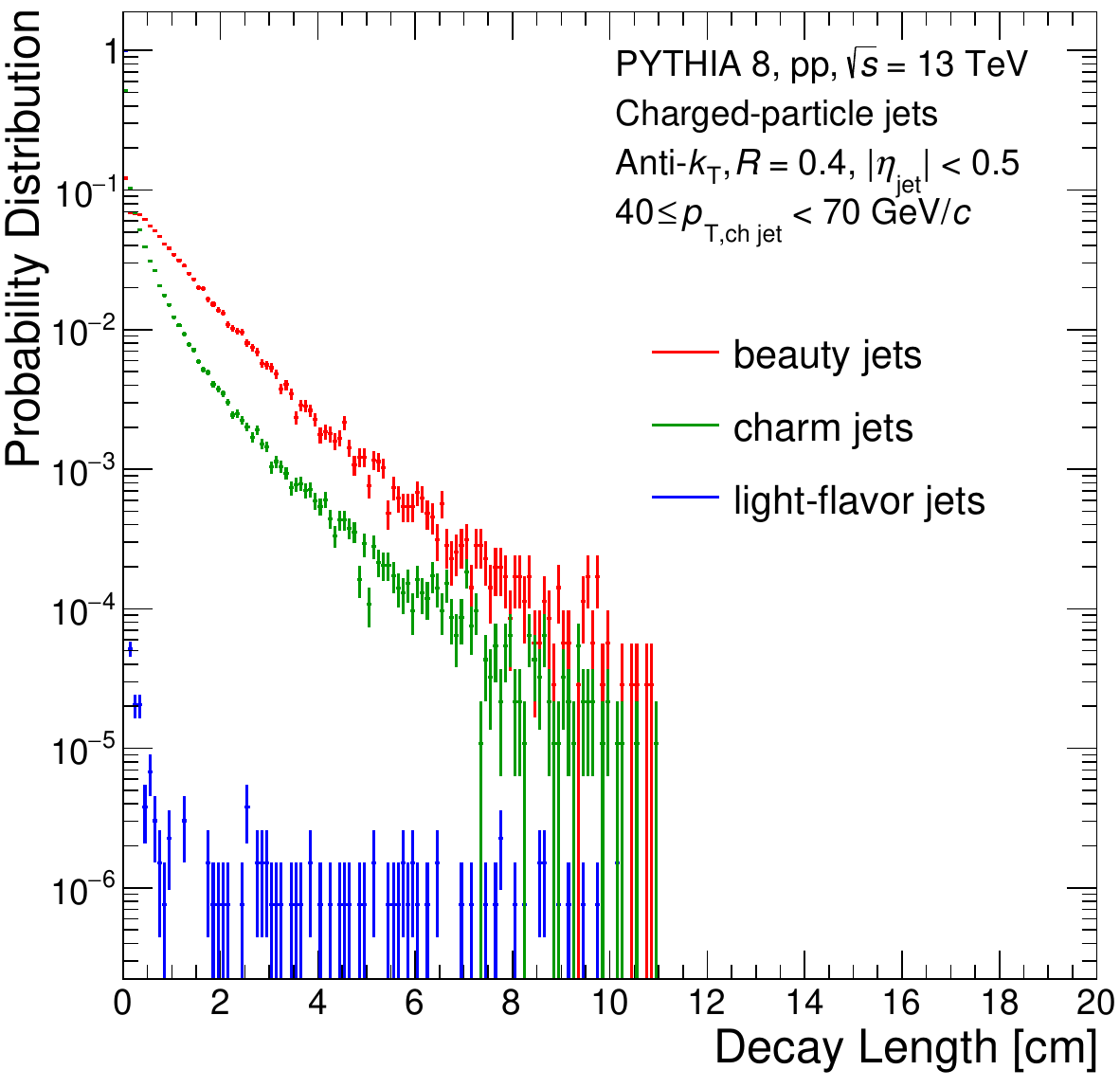}
        \label{fig:DL}
    }
    \caption{Left: Cosine pointing angle distribution of the secondary vertices for the different jet flavors. Right: Secondary vertex decay length.}
    \label{fig:SVParams}
\end{figure*}

Secondary vertices within jets are reconstructed from their constituent tracks. First, a kinematic cut is applied by selecting tracks with $p_{\rm T}>0.3$~\GeVc, which helps to exclude tracks significantly affected by detector resolution. Triplets of the selected tracks are then combined to identify potential secondary vertices under the assumption that they may originate from a common decay point. For each triplet, a vertex fitting algorithm is employed to determine the optimal vertex position by minimizing the distance between the tracks and the vertex—quantified by the $\chi^2$ fit. 
The resulting candidate vertices are subjected to additional quality criteria to ensure that only well-reconstructed and physically meaningful secondary vertices are retained. 
It should be noted here that no smearing on the secondary vertices is applied since the DCA and momentum of the tracks, which are used for the secondary vertex reconstruction, are already smeared.

The features from the reconstructed secondary vertices inside the jet are derived from the tracks associated with the vertex. These features include: SV \pt, $\eta$, the distance between the secondary vertex and the jet calculated using the $\Delta R$ equation mentioned above, the SV mass, the energy fraction carried by the secondary vertex in the jet, the distance of closest approach in the bending plane of the SV trajectory with respect to the primary vertex, the cosine pointing angle (CPA) which is the cosine of the angle between the SV momentum and the vector pointing from the primary vertex to SV, $\chi^2$ of the SV reconstruction fit, and the dispersion $D = \sqrt{(d_1^2 + d_2^2 + d_3^2)/N}$ which is the square root of squared summed DCAs of the daughter tracks with respect to the secondary vertex normalized by the number of tracks. The decay length which is the flight distance between the primary and secondary vertices in the transverse plane and in the $z$ direction is also included in the list of input features.

Figure~\ref{fig:SVParams} (right) shows the decay length distribution for the different jet flavors, and Fig.~\ref{fig:SVParams} (left) shows the cosine pointing angle distribution for the reconstructed secondary vertex. As can be seen in the figure, the decay length is very discriminative for the different jet flavors. This behavior is expected because beauty and charm hadrons have short lifetime so they decay away from the primary vertex, producing a measurable decay length. Light-flavor hadrons are produced from the primary vertex so they do not form a secondary vertex, resulting in a very small decay length. Since the decay length is the most discriminative feature for secondary vertices, all SVs inside the jets are sorted in descending order based on their decay length. The top ten SVs are then selected as input for the ML model.

The number of input tracks and SVs per jet is limited to 10. This choice is motivated by a dedicated study in which the number of input objects was varied between 3 and 20. The discrimination performance improves with increasing multiplicity up to approximately eight objects, after which it saturates, with negligible gain beyond this point. Therefore, ten inputs are used as a compromise between performance and computational cost.

Tables~\ref{tab:jet_features}, \ref{tab:track_features}, and \ref{tab:sv_features} in Appendix~\ref{sec:app_features} summarize all the features from jet, tracks, and secondary vertices and their definitions that are used for training the ML model.

\subsection{Machine learning model}\label{sec:ML}
\begin{figure*}
    \centering
    \includegraphics[width=0.98\linewidth]{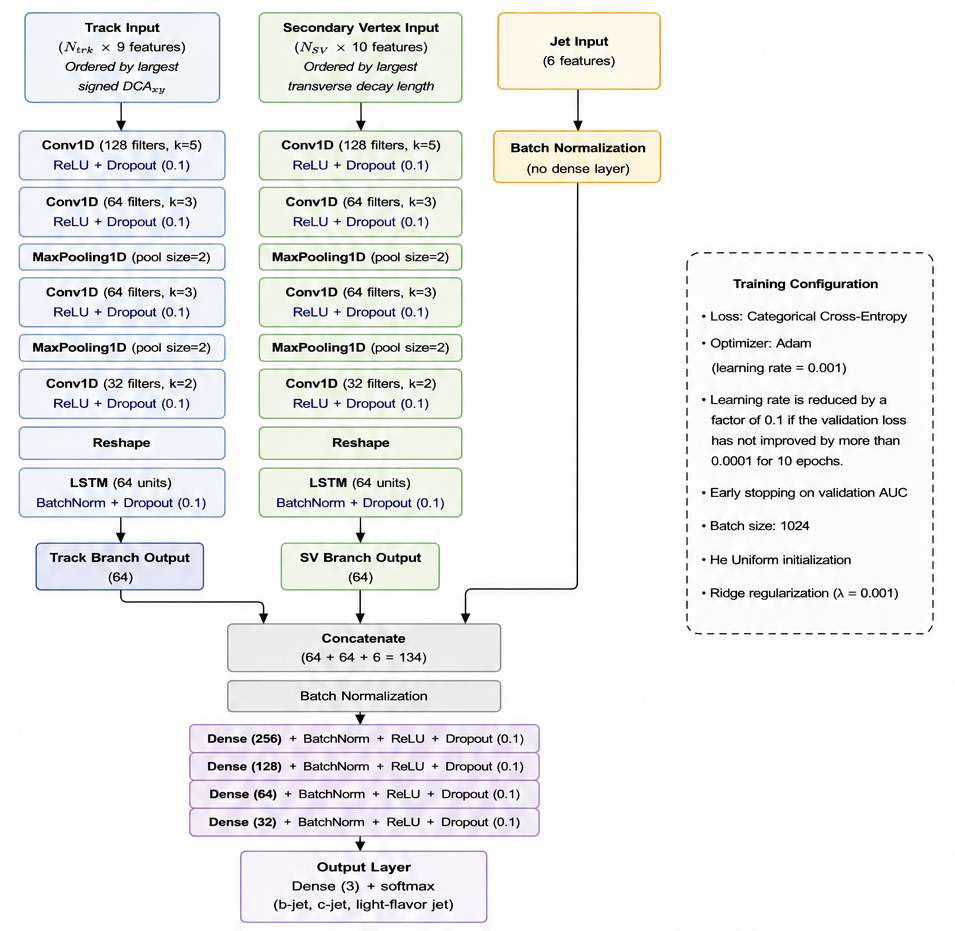}
    \caption{Machine learning model architecture combining local track and secondary vertex features processed through 1D CNNs with global jet-level features.}
    \label{fig:ModelArch}
\end{figure*}

Convolutional neural networks (CNNs)~\cite{CNN} are neural networks used specifically for image classification by identifying patterns and features in data. They consist of three types of layers: a convolution layer that applies kernel filters that extract patterns, a pooling layer that reduces the number of inputs, and fully connected layers that will make the final decision and connect the CNN to the output layer. In this analysis, a machine learning algorithm comprising three branches has been used. Two CNN branches correspond to the ten tracks and ten SVs connected at the end to a recurrent neural network (RNN) layer with long short-term memory (LSTM) units~\cite{lstm}, and a branch that corresponds to the global jet parameters.


The convolution branches consist of four convolution layers made of 128, 64, 64, and 32 filters with kernel sizes 5, 3, 3, and 2, respectively. Two max-pooling layers of size 2 are applied after the second and third convolutions. A stride of 1, same padding, and \texttt{relu} activation function are used. A dropout of 0.1 is applied after each layer. At the end of the convolution process, the output is reshaped and then connected to an LSTM layer with 64 RNN units, batch normalization, and a dropout of 0.1 at the end. The third branch, which corresponds to the jet features, will be directly concatenated with the two other branches. Then the flattened layer will be connected to four fully connected layers with 256, 128, 64, and 32 neurons. A dropout of 0.1 is again applied after each layer. The output is then connected to a softmax layer with three classes corresponding to the different jet flavors. Batch normalization is applied after each layer. The loss function used is the \texttt{categorical cross entropy} with \texttt{Adam} optimizer, He Uniform kernel initializer, ridge regularization with regularization parameter of 0.001, and a learning rate of 0.001. The learning rate is reduced by a factor of 0.1 if the validation loss has not improved by more than 0.0001 for ten epochs.
The model architecture is summarized in a diagram as shown in Fig~\ref{fig:ModelArch}. The model is a multiclass classifier that provides three output probabilities corresponding to beauty, charm, and light-flavor jets. These probabilities are combined into a single one-dimensional b-jet discriminator defined as $D_b$ \cite{ATL-PHYS-PUB-2022-027}, as follows:

\begin{equation}
    D_{b} = \log \left( \frac{P_{b}}{f_{c} P_{c} + (1 - f_{c}) P_{lf}} \right)
    \label{eq:Db}
\end{equation}
where $P_b$, $P_c$, and $P_{lf}$ are the probabilities given by the model for a jet to be classified as a b-, c-, or lf-jet, respectively, and $f_c$ is the parameter corresponding to the charm-jet fraction in the background. In this paper, the parameter $f_{c}$ was chosen to be 0.08 as expected from the inclusive jet composition in PYTHIA8.

In order to train this model, 160 million \pp collisions at \s = 13 TeV are generated using PYTHIA8. From all of these events, around $2 \times 10^{6}$ b-jets were obtained and used as input for the TensorFlow~\cite{tensorflow} ML model. The sample is constructed from three classes corresponding to b-, c-, and light-flavor jets with equal fractions. This corresponds to a prior probability of 1/3 for each class. The sample is randomly split into 70\% used for training, 15\% for validation, and 15\% for testing. Each subsample contains approximately equal fractions of b-, c-, and light-flavor jets.

After compiling the neural network with all its CNN and dense layers, it is trained on the training sample using a batch size of 1000. A maximum of 300 epochs is allowed, with early stopping applied based on the validation accuracy. Training is stopped if the validation accuracy has not improved by more than $10^{-4}$ for ten consecutive epochs. In practice, convergence is reached much earlier, and the optimal model is selected after 53 epochs, thereby preventing overfitting. The performance is then evaluated on the testing sample. Finally, the trained network is converted into ONNX~\cite{onnx} format, which is loaded and applied to the reconstructed jets during event generation.

\section{Results}\label{sec:results}
\begin{figure}[htb]
    \centering
    \subfloat{
        \includegraphics[width=0.75\linewidth]{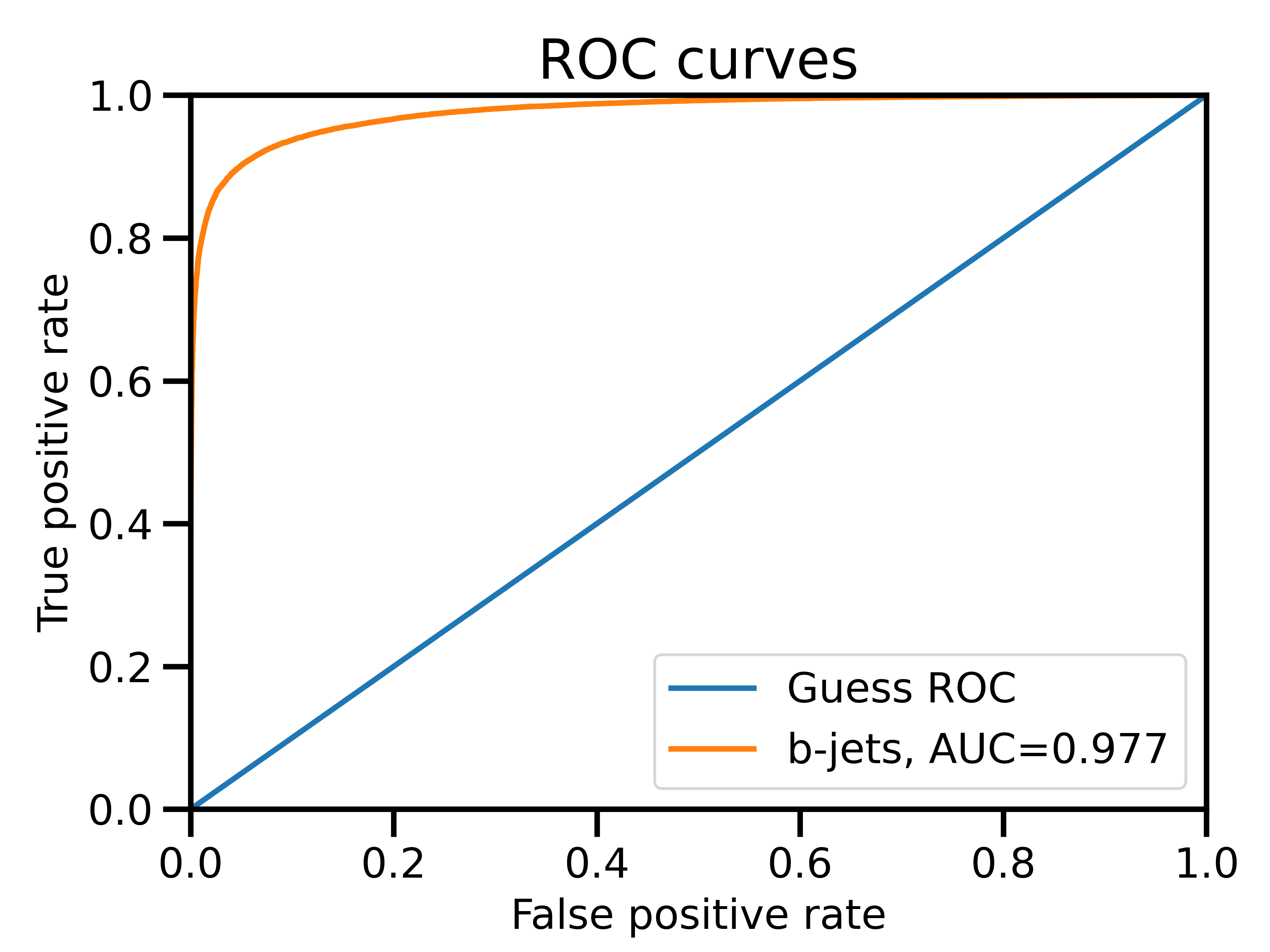}
        \label{fig:ROC}
    }
    \hfill
    \subfloat{
        \includegraphics[width=0.75\linewidth]{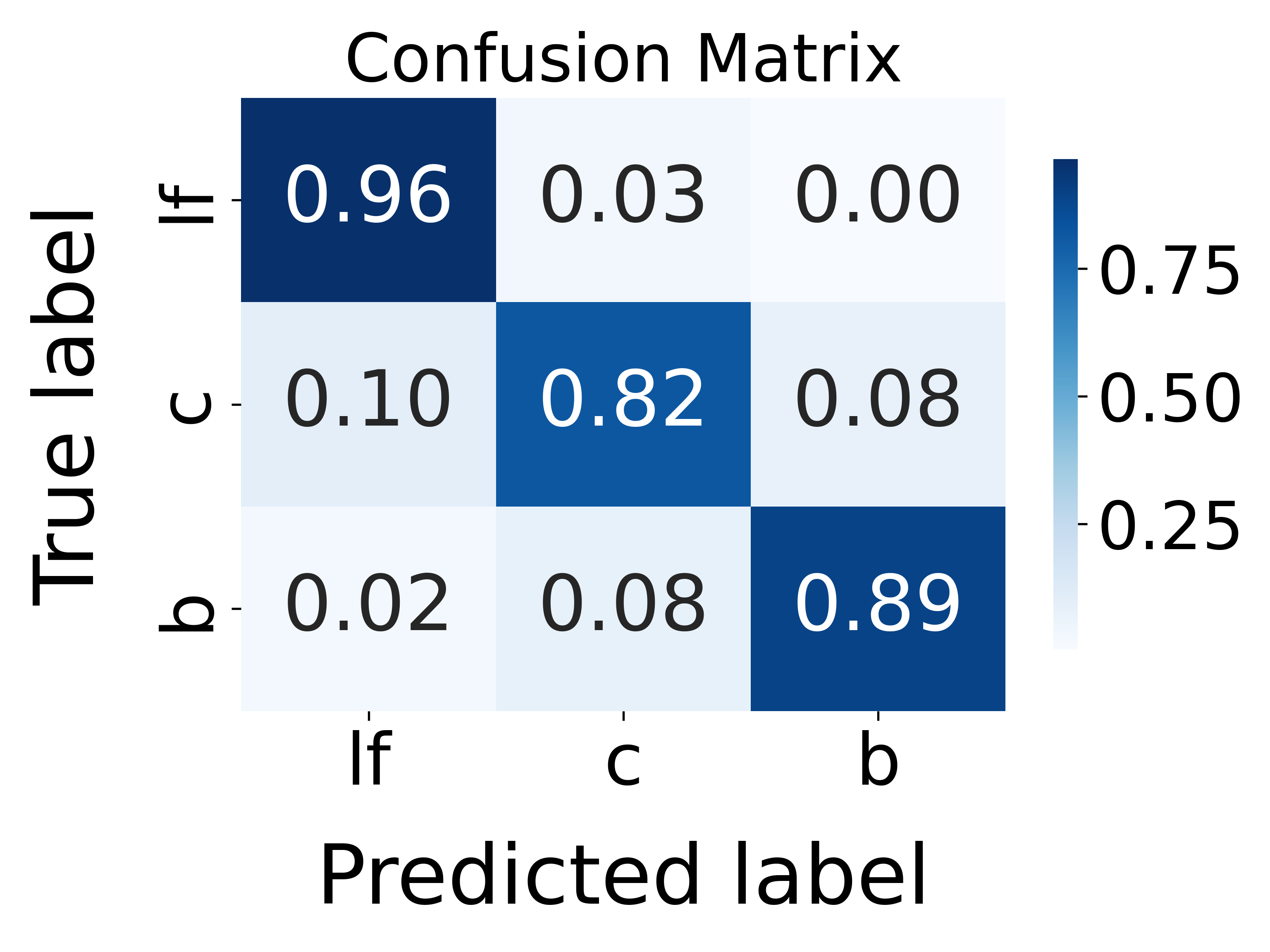}
        \label{fig:Conf}
    }
    \caption{ROC curves and confusion matrix}
    \label{fig:ROC_Conf}
\end{figure}

Figure~\ref{fig:ROC_Conf} (top) shows the receiver operating characteristic (ROC) curve of the ML algorithm. The ROC curve is a graphical representation for the model performance. It shows the true-positive rate (tagged b-jets) versus the false positive rate (mistagged c-/lf-jets) at various thresholds. As can be seen in the figure, there is a small decrease in the true-positive rate compared to a sharp decrease in false-positive rate. This figure also shows the area under the curve (AUC), which is the area under the ROC curve; if the AUC is 1.0, it means perfect separation between signal and background, while 0.5 means random guessing. The AUC shown in Fig.~\ref{fig:ROC_Conf} (top) is around 0.977, which shows that the model is able to efficiently separate the different jet flavors.
The whole training and evaluation procedures were repeated five times using different random seeds for splitting the data between train/validation/test in order to test the model stability. By comparing the ROC curves obtained from these different trainings, the results were found to be consistent with an average AUC of $0.9754 \pm 0.0012$.

Figure~\ref{fig:ROC_Conf} (bottom) shows the confusion matrix, which is a table that summarizes the performance of the model by comparing the true labels of jets to the predicted ones from the ML model. It displays the fraction of correctly classified b-, c-, and lf-jets along the diagonal entries, and the fraction of misidentified jets elsewhere. This misidentification can happen either due to the presence of secondaries in the lf-jets or due to the presence of a displaced secondary vertex in the case of charm jets.
As shown in the figure, the fraction of misidentification is very small compared to the fraction of correctly identified jets, which again reflects the high purity of the tagged jets sample. The b-jets are tagged with an accuracy of almost 90\%, which is a decent score based on the complexity of the problem.

\begin{figure}[htb]
    \centering
    \subfloat{
        \includegraphics[width=0.75\linewidth]{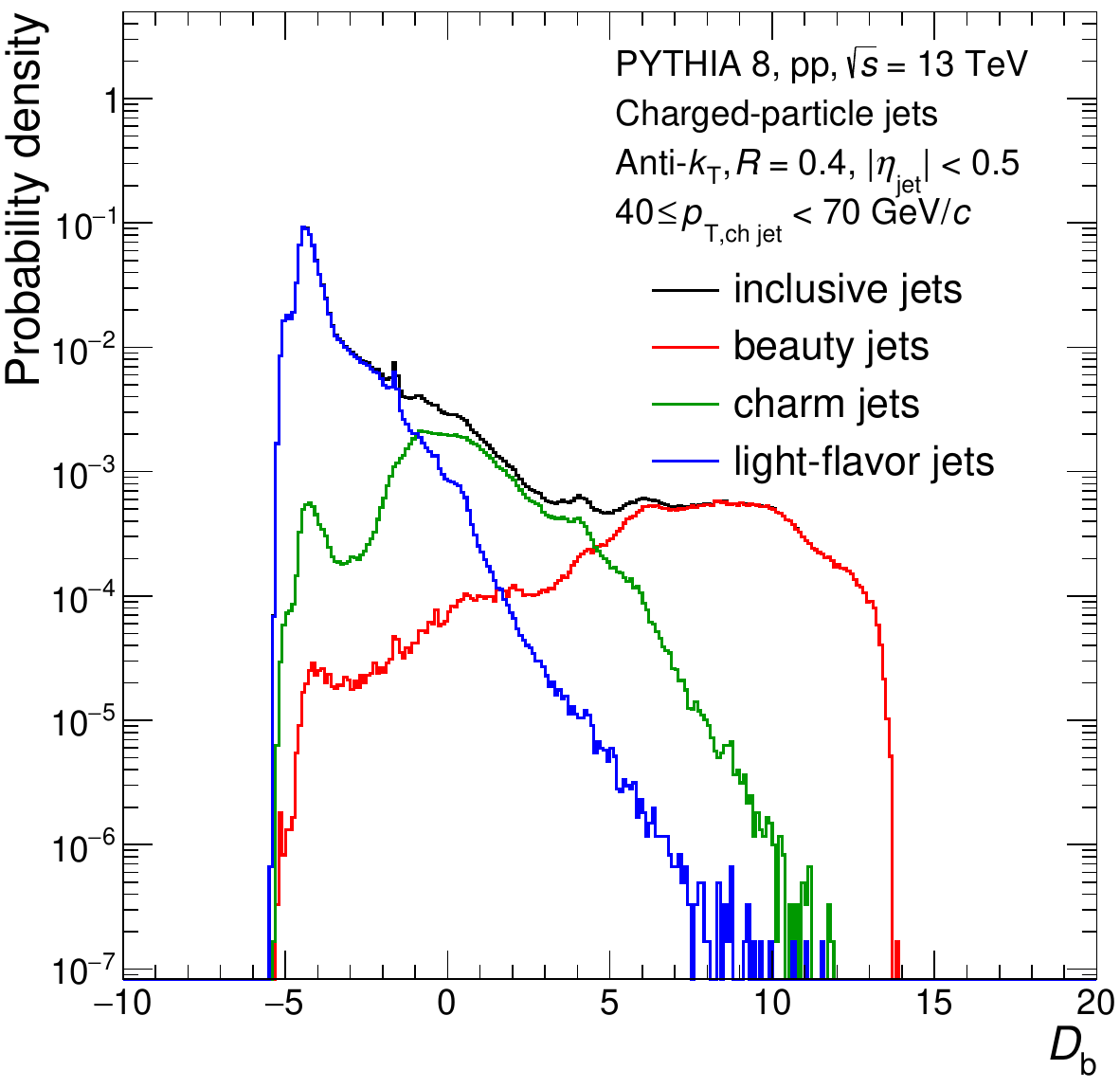}
        \label{fig:ScoreDist}
    }
    \hfill
    \subfloat{
        \includegraphics[width=0.75\linewidth]{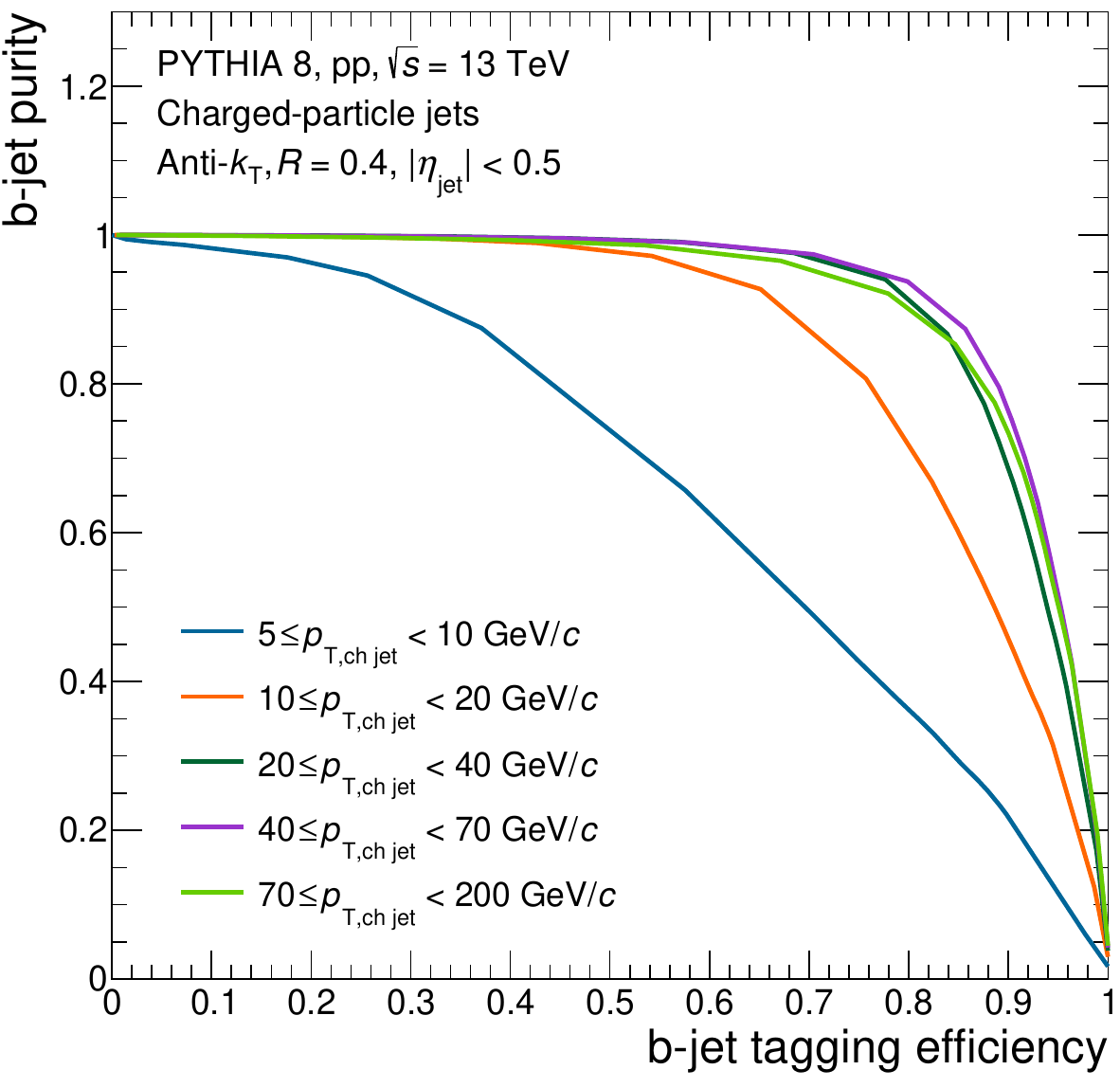}
        \label{fig:PurityVsEfficiency}
    }
    \caption{Top: $D_b$ distribution for the different jet flavors. Bottom: b-jet purity as a function of the tagging efficiency from the ML model.}
    \label{fig:PurityEff}
\end{figure}

Figure~\ref{fig:PurityEff} (top) shows the $D_b$ distribution of the different jet flavors as defined in Eq.~\eqref{eq:Db}.
As one can observe, the distribution decreases sharply for the light-flavor jets. For charm jets, the distribution increases slightly to a peak at $D_b = 0$ before experiencing a sharp decrease. For beauty jets, however, the distribution gradually increases and reaches a peak around $D_b = 6$ before decreasing at higher values. This demonstrates the model's separation ability between b-jets and c-/lf-jets.

\begin{figure*}[!htbp]
    \centering
    \subfloat{
        \includegraphics[width=0.45\linewidth]{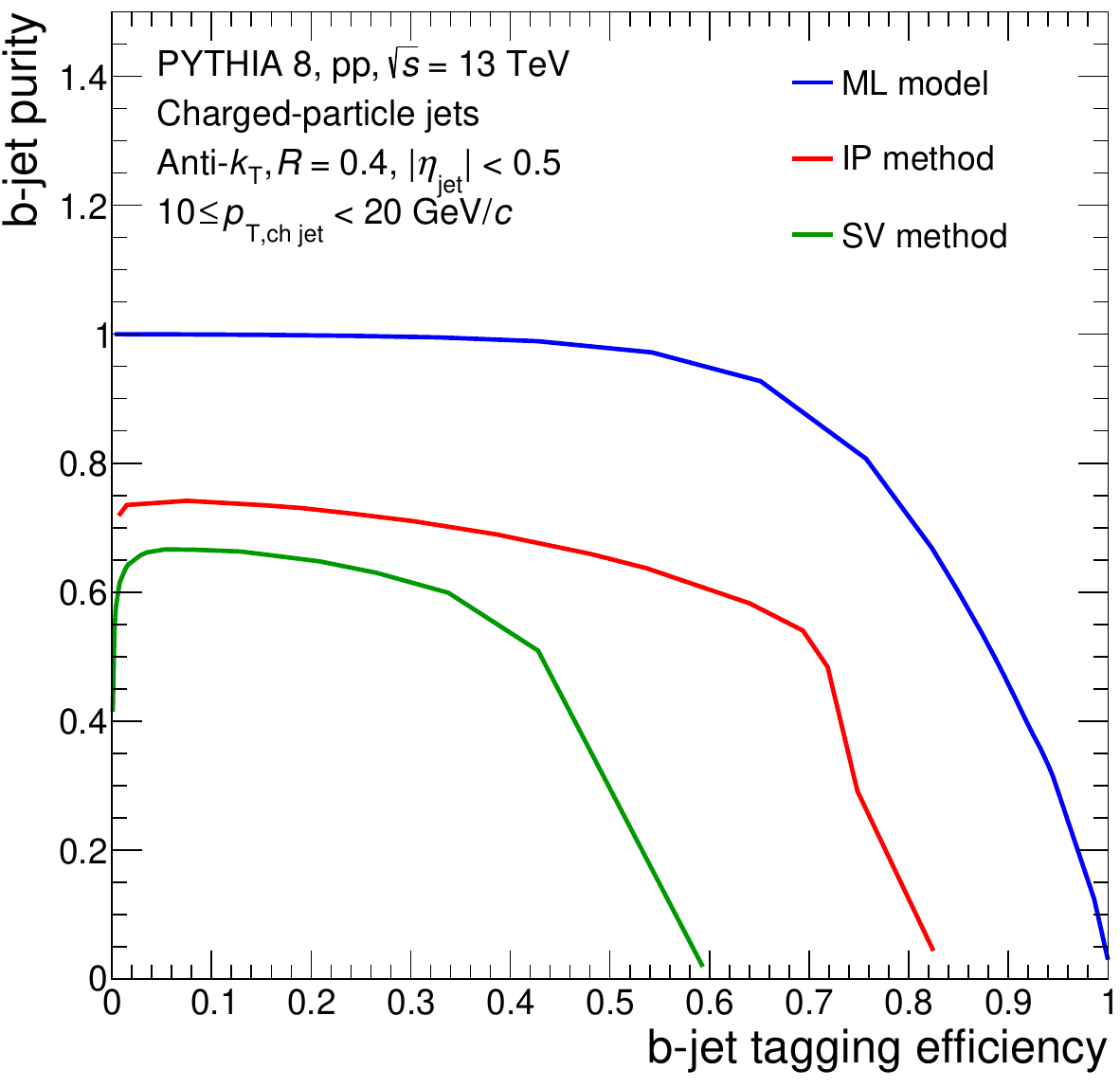}
        \label{fig:PurEff_10_20}
    }
    \subfloat{
        \includegraphics[width=0.45\linewidth]{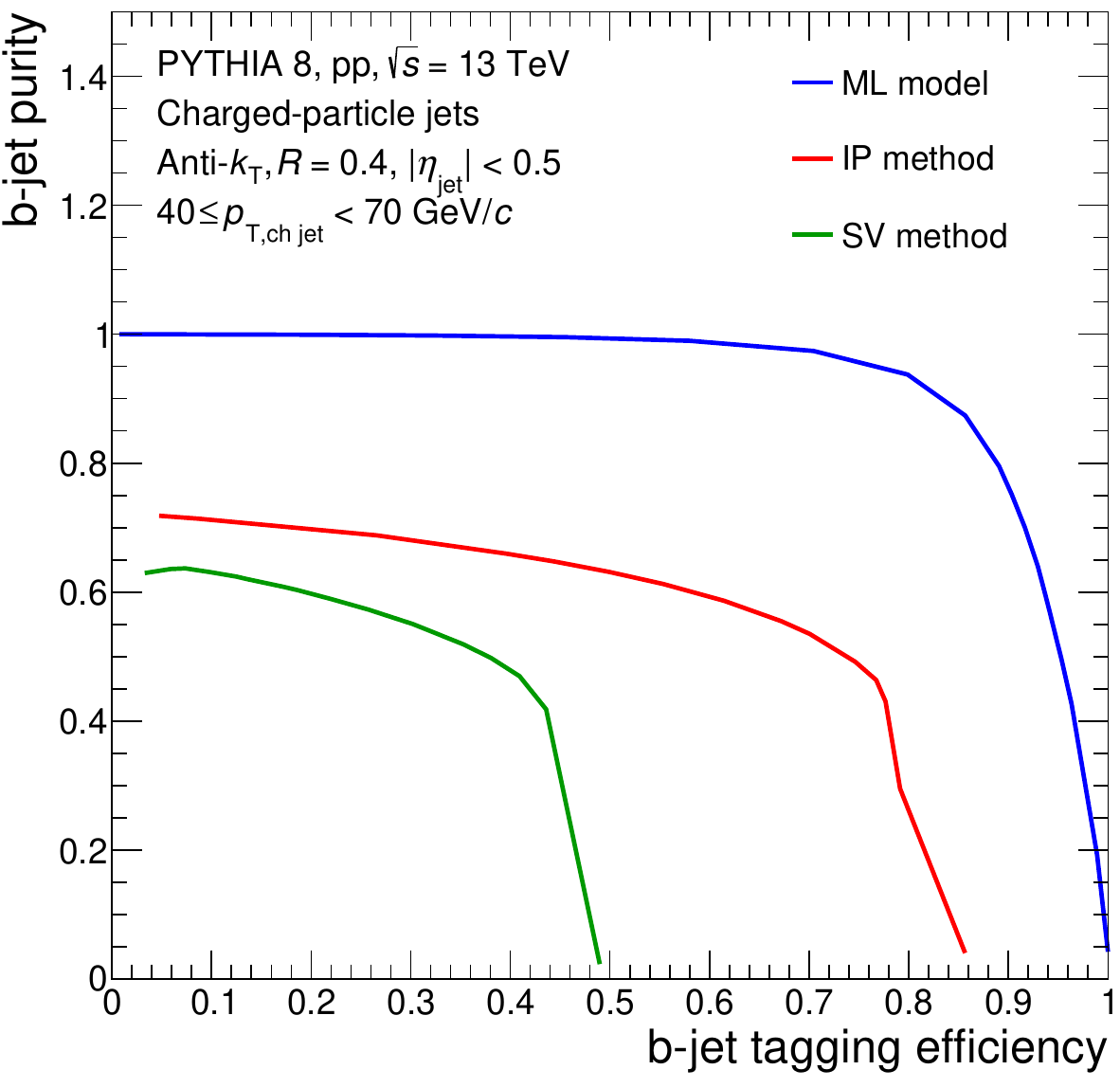}
        \label{fig:PurEff_40_70}
    }
    \caption{ML model performance (purity versus efficiency) compared with the performance of traditional methods (IP and SV methods) in two different \pt ranges.}
    \label{fig:Tradional}
\end{figure*}

Figure~\ref{fig:PurityEff} (bottom) shows the purity versus efficiency of the b-tagging model for different jet \pt bins. This figure is crucial since it illustrates the performance of the model in different kinematic regions. It is made by changing the $D_b$ threshold from -5 to 13 and selecting all the jets that pass this threshold. 
As shown in the figure, low score threshold values will give high efficiencies but low purity, since this allows for more background jets to be tagged. Conversely, the high score threshold values lead to lower efficiencies but, in return, give a very high purity since it suppresses a large fraction of the background. High efficiency might be needed in measurements that require a maximum signal yield, while purity is crucial for analyses where the background contamination must be minimized, such as when measuring b-jets in \PbPb collisions at LHC energies.

When comparing the model performance in the different \pt regions, one can see that the performance of the ML model is reliable (it reaches 90$\%$ purity at $80\%$ efficiency) for high \pt jets, which is due to the larger DCA of the tracks and the larger decay length of the secondary vertices compared to charm and light-flavor jets. The performance at the lowest \pt bins is not as good as the high \pt jets, which is due to the low particle multiplicity inside the jet compared to the high \pt jets. This is also due to the shorter decay lengths which might make the jet resemble charm or light-flavor jet, imposing higher complexity for the jet flavor tagging at low-$p_{\rm T}$.

Finally, to validate the use of machine learning–based models for jet flavor tagging, Fig.\ref{fig:Tradional} shows the performance of the ML model compared to traditional cut-based methods, such as the secondary vertex and track counting methods. These approaches rely on the track with the largest DCA in the jet and are implemented following the procedure described in Ref.~\cite{ALICE:2021wct}. As can be seen, the ML model used in this work achieves a much higher purity for the same efficiencies. For example, at 50$\%$ efficiency, the ML model achieves $30\%$ higher purity compared to the IP method. This means that the ML model is superior since other methods rely on single observables such as the SV decay length or DCA, while the ML model, working on a multivariate analysis framework, fuses all the possible observables from the jet, tracks, and secondary vertices. This demonstrates the advantage of using machine learning models in jet flavor tagging.

\section{Conclusion}\label{sec:conclusion}

Events in \pp collisions at \s = 13 TeV are generated using PYTHIA8 (Monash) model, where charged particles with $\pt > $ 0.15 \GeVc and $|\eta| < $ 0.9 are selected. The \pt and DCA of the selected particles are smeared according to the ALICE ITS2 detector transverse momentum and DCA resolutions, respectively. Then, jets are reconstructed from these charged particles using the anti-$k_{T}$ algorithm with $R=$ 0.4. Then a machine learning model that uses CNNs is trained and tested on these jets. The result shows a better performance in terms of purity and tagging efficiency of the ML model compared to the traditional methods.

The current study is performed only on events generated using PYTHIA8 simulation. This model is planned to be used on the ALICE experimental data in the future, which requires a good agreement between data and MC for the input variables and discriminator output ($D_b$). Possible discrepancies between data and MC could affect the performance of the model and may require further investigation in the future.

~~~~~~~~~~
\begin{acknowledgments}
We would like to thank the Jyväskylä ALICE group for fruitful discussions.
We acknowledge CSC-IT Center for Science in Espoo, Finland, for allocating computational resources.
N.M. and D.J.K. are supported by the Academy of Finland through the Center of Excellence in Quark Matter (Grant No. 346328), and H.H. through Grant No. 346327. H.H. also acknowledges support from JSPS KAKENHI Grant No. JP24H00003.
\end{acknowledgments}

~~~~~~~~~~
\appendix

\section{The features used for training the ML model}\label{sec:app_features}

\begin{table*}[!htbp]
\centering
\caption{List of jet-level features used as input for the ML model.}
\begin{tabular}{lp{0.6\linewidth}}
\hline
\textbf{Feature} & \textbf{Description} \\
\hline
Jet $p_{\mathrm{T}}$ & Transverse momentum of the jet \\
Jet $\eta$ & Pseudorapidity of the jet \\
Jet $\varphi$ & Azimuthal angle of the jet \\
Number of track constituents & Number of tracks associated with the jet \\
Number of secondary vertices & Number of reconstructed SVs inside the jet \\
Jet mass & Invariant mass of the jet \\
\hline
\end{tabular}
\label{tab:jet_features}
\end{table*}

\begin{table*}[!htbp]
\centering
\caption{List of jet constituent (track-level) features used as input for the ML model.}
\begin{tabular}{lp{0.65\linewidth}}
\hline
\textbf{Feature} & \textbf{Description} \\
\hline
Particle $p_{\mathrm{T}}$ & Transverse momentum of the particle \\
Particle $\eta$ & Pseudorapidity of the particle \\
$\vec{p}_{\mathrm{jet}} \cdot \vec{p}_{\mathrm{part}}$ & Dot product of jet and particle momenta \\
$\vec{p}_{\mathrm{jet}} \cdot \vec{p}_{\mathrm{part}}/|\vec{p}_{\mathrm{jet}}|$ & Normalized dot product by jet momentum \\
$\Delta R$ (jet, track) & Distance between jet axis and track in $\eta$–$\varphi$ plane \\
Signed DCA ($xy$) & Signed distance of closest approach in transverse plane \\
Signed DCA ($z$) & Signed distance of closest approach in $z$ direction \\
Distance to closest SV & Distance between track and closest SV \\
\hline
\end{tabular}
\label{tab:track_features}
\end{table*}

\begin{table*}[!htbp]
\centering
\caption{List of secondary vertex features used as input for the ML model.}
\begin{tabular}{lp{0.6\linewidth}}
\hline
\textbf{Feature} & \textbf{Description} \\
\hline
SV $p_{\mathrm{T}}$ & Transverse momentum of the SV \\
SV $\eta$ & Pseudorapidity of the SV \\
$\Delta R$(SV, jet) & Distance between SV and jet using $\Delta R$ \\
SV mass & Invariant mass of the SV \\
Energy fraction & Fraction of jet energy carried by the SV \\
DCA (bending plane) & Closest approach of SV trajectory to the primary vertex \\
CPA & Cosine of angle between SV momentum and the pointing vector PV $\to$ SV \\
$\chi^2$ of SV fit & $\chi^2$ of the SV reconstruction fit \\
Dispersion $D$ & $\sqrt{(d_1^2 + d_2^2 + d_3^2)/N}$; summed squared DCAs of daughter tracks divided by $N$ \\
Decay length (transverse, $z$) & Flight distance between PV and SV in transverse plane and $z$ \\
\hline
\end{tabular}
\label{tab:sv_features}
\end{table*}

\newpage
\bibliographystyle{utphys}   
\bibliography{main}

\end{document}